\documentstyle[11pt,newpasp,twoside,psfig]{article}
\index{instructions}
\index{guidelines}

\def\edcomment#1{\iffalse\marginpar{\raggedright\sl#1\/}\else\relax\fi}
\reversemarginpar

\begin{document}
\title{The ATNF Pulsar Catalogue}
 \author{G. Hobbs, R. Manchester, A. Teoh, M. Hobbs}
\affil{Australia Telescope National Facility -- CSIRO, P.O. Box 76,
 Epping NSW 1710, Australia}

\begin{abstract}
The number of known pulsars has significantly increased over the
previous few years. We have searched the literature to
find papers announcing the discovery of pulsars or giving improved
parameters for them.  Data from the papers have been entered into a
new pulsar catalogue that can be accessed via a web interface or
from the command line (on Solaris or Linux machines).  The user may
request over 120 different parameters, select pulsars of interest,
generate custom variables and choose between different ways of
displaying or tabulating the data.  Full bibliographic references are
available on all observed parameters.
\end{abstract}

\section{Introduction}

We have produced a new catalogue (accessed via a web or
command-line interface) that contains all published rotation--powered
pulsars, including those detected only at high energies.  The
catalogue also contains anomalous X-ray pulsars (AXPs) and soft
gamma-ray repeaters (SGRs) for which coherent pulsations have been
detected.  We have excluded accretion--powered pulsars such as Her X-1
and the recently discovered X-ray millisecond pulsars (e.g. SAX
J1808.4$-$3658; Wijnands \& van der Klis 1998).  We currently store
information on 1300 pulsars.

For each pulsar we record up to 90 different parameters with full
bibliographic information.  For all pulsars we store names, positions,
rotational frequencies and the type of pulsar (e.g. AXP).  If
published values exist, we also store parameters such as higher-order
rotational frequency derivatives, pulse widths, orbital parameters,
proper motions and any associations with globular clusters, the
Magellanic clouds or supernova remnants.  If enough information exists
to calculate derived parameters then these may
also be displayed using the web or command-line interfaces. Currently, 39
different derived parameters are included.

\section{The web interface}

A versatile web interface ({\tt
http://www.atnf.csiro.au/research/pulsar/psrcat}) has been developed
to allow easy global access to our pulsar catalogue.  The user may
select parameters of interest, generate custom variables, select
wanted (or unwanted) pulsars and tabulate the results in a variety of
different forms.  For browsers with Java enabled, the results may also
be displayed graphically, as a two dimensional plot or as a histogram,
using an interactive applet (Figure 1).
We also provide a special `expert-mode' of operation which can be
accessed from {\tt
http://www.atnf.csiro.au/research/pulsar/psrcat/expert.html}. This mode
allows the user to obtain pulsar parameters of specialist
interest and to upload one or more personal
catalogues to use instead of, or to merge with, the public
catalogue. Such merging is not permanent and does not affect any other
user of the system.
\begin{figure}
\begin{center}
\psfig{file=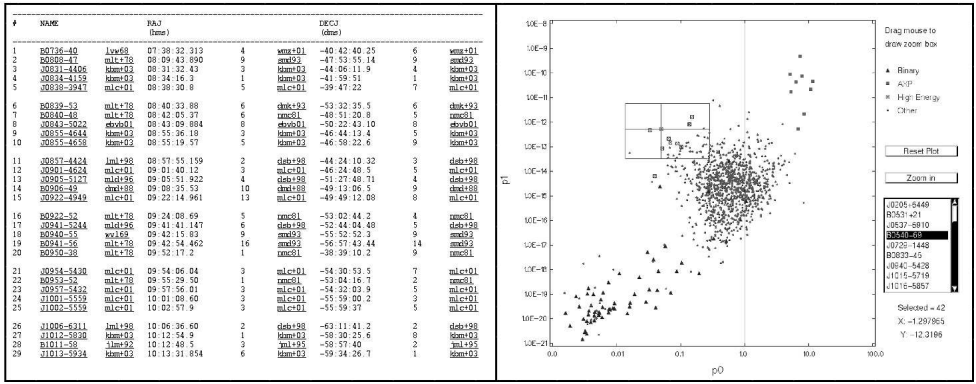,width=13cm}
\caption{An example of the `long-form' of tabular output and the
graphical plotting interface.}
\end{center}
\end{figure}

\section{The command-line interface}

Data can be obtained from the public or from a personal catalogue
using a command line interface.  This software has been tested on both
Solaris and Linux machines and can be obtained via ATNF anonymous ftp
(see authors for details). This software provides the same
functionality as the web-based interface with the exception of the
plotting routines.  C functions are available to access the catalogue
data directly.

\section{Acknowledgements}
Many people have contributed to the maintenance and upgrading of
earlier versions of the database. We particularly thank Andrew Lyne of
the University of Manchester, David Nice of Princeton University and
Russell Edwards, then at Swinburne University of Technology.  This
work has made extensive use of NASA's Astrophysics Data system and the SLALIB C routines.

\end{document}